\documentstyle[12pt,eqsecnum,aps,prb,tighten]{revtex}

\newcommand{\be}{\begin{equation}}
\newcommand{\ee}{\end{equation}}

\newcommand{\gapproxeq}{{\ 
\lower-1.2pt\vbox{\hbox{\rlap{$>$}\lower5pt\vbox{\hbox{$\sim$}}}}\ }}
\newcommand{\lapproxeq}{{\ 
\lower-1.2pt\vbox{\hbox{\rlap{$<$}\lower5pt\vbox{\hbox{$\sim$}}}}\ }}
\newcommand{\1}[1]{\stackrel{\tiny{1}}{#1}}
\newcommand{\2}[1]{\stackrel{\tiny{2}}{#1}}

\begin{document}
\setlength{\unitlength}{1mm}

\title{Scattering of Straight Cosmic Strings by \\
Black Holes:  Weak Field Approximation
\footnote{Preprint Alberta Thy 04-98}}

\vspace{1cm}

\author{
Jean-Pierre De Villiers\footnote{e-mail:
jpd@phys.ualberta.ca}  ${}^{1}$, and
Valeri Frolov\footnote{e-mail: frolov@phys.ualberta.ca} ${}^{1,2}$\\
\vspace{1.5cm}}
\maketitle
${}^{1}${\em
Theoretical Physics Institute, Department of Physics, 
University of
Alberta, \\   Edmonton, Canada T6G 2J1}
\\${}^{2}${\em CIAR Cosmology Program}

\bigskip
\vspace{12 pt}

\begin{abstract}  
\noindent 
The scattering of a straight, infinitely long string moving with
velocity $v$ by a black hole is considered. We analyze the weak-field
case, where the impact parameter ($b_{imp}$) is large, and obtain exact
solutions to the equations of motion. As a result of scattering, the
string is displaced  in the direction perpendicular to the velocity by
an amount  $\Delta b\sim -2\pi GMv\gamma/c^3 -\pi (GM)^2/ (4c^3 v
b_{imp})$, where $\gamma=(1-(v/c)^2)^{-1/2}$.
The second term dominates at low velocities
$v/c<(GM/b_{imp})^{1/2}$ . The late-time solution is represented by a
kink and anti-kink, propagating in opposite directions at the speed of
light, and leaving behind them the string in a new ``phase''. The
solutions are applied to the problem of string capture, and are
compared to numerical results. 
\end{abstract}

\vspace{3cm}

{\it PACS number(s): 04.60.+n, 12.25.+e, 97.60.Lf, 11.10.Gh}

\newpage
\baselineskip=.8cm

\section{Introduction}

A cosmic string is a relativistic non-local object with an infinite
number of internal degrees of freedom. The problem of scattering and
capture of a cosmic string by a black hole is interesting for many
reasons. In some regimes, it has features in common with the scattering
of test particles. In other regimes, its non-local properties give rise
to similarities with the problem of black-hole--black-hole scattering.
In the process of scattering or capture, one can expect strong
gravitational radiation from the string-black hole system; this
radiation might be of astrophysical interest in connection with LIGO
and other projects searching for gravitational radiation. 

In our study of string motion we neglect the gravitational effects
produced by the string (which for GUT strings are of order $\sim
10^{-6}$) and assume that the width of the string is negligible (for
GUT strings the width is of order $\sim 10^{-29}$ cm). In this
approximation, a test cosmic string is represented by a two-dimensional
world-sheet, and its motion is described by the Nambu-Goto action
\cite{ShVi:94}.  From the mathematical point of view, the scattering
problem reduces to finding a minimal surface which gives an extremum to
the Nambu-Goto action. 

We are interested in a cosmic string whose length is much greater than
the radius of the black hole.  For this reason, we will consider a
string of infinite length. The interaction of the string with a black
hole has two possible outcomes: either the string is captured by the
black hole, or it is scattered. In the latter case the string absorbs
some energy, so this process is inelastic. 

A complete description of the final stationary configurations of
trapped cosmic strings has already been given
\cite{FrHeLa:96,FrHeDV:97}. Stationary trapped strings are a special
case of stationary string configurations; in the Kerr-Newman spacetime,
stationary string configurations admit exact solutions  by separation
of variables \cite{FrSkZeHe:89,CaFr:89}. The general scattering
problem, and the determination of the conditions of capture, requires
solving the dynamical equations and is a much more complicated problem.
A numerical determination of the critical impact parameter for capture
has been discussed in Ref.\cite{LoMo:88,DVFr:97}. 

This paper is devoted to the analytical study of the motion of a
straight cosmic string in the gravitational field of a black hole in
the weak-field approximation.  At early times (before scattering), and
at late times (after scattering), the string is moving in a nearly flat
spacetime where the weak-field approximation allows one to formulate
the scattering problem in terms of ``in'' and ``out'' states of the
string. For large impact parameters, the string moves at all times in a
region where the weak-field approximation remains valid. Moreover, even
if the impact parameter is small and the string reaches the
strong-field region near the black hole, the analytic weak-field 
solutions of the equations of motion are important in formulating the
initial and boundary conditions for the numerical computations
\cite{DVFr:98}. 

In this paper we derive and solve the equations of motion of an
infinite straight cosmic passing near a black hole in the weak-field
approximation. We demonstrate that, as a result of scattering, the
string is displaced  in the direction perpendicular to its motion by an
amount  $\Delta b\sim -2\pi GMv\gamma/c^3 -\pi (GM)^2/ (4c^3 v
b_{imp})$, where $\gamma=(1-(v/c)^2)^{-1/2}$.
The second term dominates at low velocities
$v/c<(GM/b_{imp})^{1/2}$.  This
result for low velocity motion is in an agreement with the result
recently obtained by Page \cite{Page:98}. The late-time solution is
represented by a kink and anti-kink, propagating in opposite directions
at the speed of light, and leaving behind them the string in a new
``phase''. In the Conclusions, the solutions are applied to the problem
of string capture, and are compared to numerical results.

\section{Motion of Straight Strings}

The aim of this paper is to study the scattering of an infinitely long
cosmic string by a black hole. We assume that the string is initially
far from the black hole, straight, and moving with constant velocity
$v$. We assume that the gravitational field is weak and solve the
equations of string motion using the perturbation theory.

Our starting point is the Polyakov action for the relativistic string
\cite{Poly:81},
\begin{equation}\label{n1.1} 
I=-{\mu \over 2}\,\int d\tau d\sigma \sqrt{-h}h^{AB}G_{AB}\, .
\end{equation}
We use units in which $G=c=1$, and the sign conventions of \cite{MTW}. In 
(\ref{n1.1}) $h_{AB}$ is the internal metric
with determinant $h$, and $G_{AB}$ is the induced metric on the
world-sheet,
\begin{equation}\label{n1.2} 
G_{AB}=g_{\mu\nu}{\partial {\cal X}^{\mu}\over \partial\zeta^A}{\partial
{\cal X}^{\nu}\over \partial\zeta^B}=g_{\mu\nu} {\cal X}^{\mu}_{,A}{\cal
X}^{\nu}_{,B} \, .
\end{equation}
${\cal X}^{\mu}$ ($\mu=0,1,2,3$) are the spacetime coordinates and
$\zeta^A$ ($A=0,3$) are the world-sheet coordinates $\zeta^0=\tau$,
$\zeta^3=\sigma$. Finally, $g_{\mu\nu}$ is the spacetime metric. 

The variation of the action (\ref{n1.1}) with respect to ${\cal X}^{\mu}$
and $h_{AB}$ gives the following equations of motion:
\begin{equation}\label{n1.3} 
\Box {\cal X}^{\mu}+h^{AB}\Gamma^{\mu}_{\alpha\beta}{\cal
X}^{\alpha}_{,A}{\cal X}^{\beta}_{,B}=0\, ,
\end{equation}
\begin{equation}\label{n1.4} 
G_{AB}-{1\over 2}h_{AB}h^{CD}G_{CD}=0 \, ,
\end{equation}
where
\begin{equation}\label{n1.5} 
\Box ={1\over \sqrt{-h}}\partial_A(\sqrt{-h}h^{AB}\partial_B)\, .
\end{equation}
The first of these equations is the dynamical equation for string
motion, while the second one plays the role of  constraints. 

In the absence of the external gravitational field
$g_{\mu\nu}=\eta_{\mu\nu}$, where $\eta_{\mu\nu}$ is the flat spacetime
metric. In Cartesian coordinates ($T,X,Y,Z$),
$\eta_{\mu\nu}=\mbox{diag}(-1,1,1,1)$ and
$\Gamma^{\mu}_{\alpha\beta}=0$, and it is easy to verify that
\begin{equation}\label{n1.6} 
{\cal X}^{\mu}=X^{\mu}(\tau,\sigma)\equiv (\cosh(\beta)\, \tau,
(\sinh\beta)\,\tau+X_0, Y_0,\sigma) \, ,
\end{equation}
\begin{equation}\label{n1.7} 
h_{AB}=\eta_{AB}\equiv \mbox{diag}(-1,1)\, ,
\end{equation}
is a solution of equations (\ref{n1.3}) and (\ref{n1.4}). This solution
describes a straight string oriented along the $Z$-axis which moves in
the $X$-direction with constant velocity $v=\tanh \beta$. Initially, at
${\tau}_{0} = 0$, the string is found at  ${\cal X}^{\mu}(0,\sigma) =
(0,X_0, Y_0,\sigma)$, with $Y_0$ playing the role of impact parameter,
$Y_0 \equiv {b}_{imp}$. For definiteness we choose $Y_0>0$ and $X_0<0$,
so that $\beta>0$.

It is convenient to introduce an orthogonal tetrad $e^{\mu}_{(m)}$
($m=0,1,2,3$) connected with the world-plane of the string
\begin{equation}\label{n1.8} 
e_{(0)}^{\mu}=X_{,\tau}=(\cosh\beta, \sinh\beta, 0, 0) \, ,
\hspace{0.5cm}e_{(3)}^{\mu}=X_{,\sigma}=(0, 0, 0, 1) \, ,
\end{equation}
\begin{equation}\label{n1.9}
e_{(1)}^{\mu}=n_1^{\mu}=(\sinh\beta, \cosh\beta, 0, 0) \, ,
\hspace{0.5cm}e_{(2)}^{\mu}=n_2^{\mu}=(0, 0, 1, 0) \, .
\end{equation}
The first two unit vectors $X_{,A}^{\mu}$ are tangent to the world-sheet
of the string, while the other two $n_R^{\mu}$ ($R=1,2$) are orthogonal 
to it.
It is easy to very that the induced metric $G_{AB}$ on the world-sheet
of the string is of the form
\begin{equation}\label{n1.10} 
G_{AB}=\stackrel{0}{G}_{AB}=\eta_{AB}\, .
\end{equation}

\section{Weak-Field Approximation}

The unperturbed solution is expressed in Cartesian coordinates. To treat
the Schwarzschild black hole as a source of perturbations on a flat
background, we  use isotropic coordinates $(T,X,Y,Z)$ in which
the line element of Schwarzschild spacetime is
\begin{equation}\label{n2.1} 
d{s}^{2}= 
-{{\left(1 - M/2R \right)}^{2}\over {\left(1 + M/2R
\right)}^{2}}\,d{T}^{2} +
{\left(1 + {M \over 2\,R}
\right)}^{4}\,\left(d{X}^{2}+d{Y}^{2}+d{Z}^{2}\right)\, ,
\end{equation} 
where $R^2=X^2+Y^2+Z^2$. This metric is of the form
\begin{equation}\label{n2.2} 
ds^2=-(1-2\Phi)dT^2+(1+2\Psi)(dX^2+dY^2+dZ^2)\, 
\end{equation}
with
\begin{equation}\label{n2.3} 
\Phi={\varphi\over (1+{1\over 2}\varphi)^2}\, ,\hspace{0.5cm}
\Psi=\varphi+{3\over 4}\varphi^2+{1\over 4}\varphi^3 +
{1\over 32}\varphi^4\,,
\end{equation}
and $\varphi$ is the Newtonian potential $\varphi= M/R$. 

In what follows we  assume that this potential is small and
write\footnote{
The same form (\ref{n2.2}) of the metric is
valid for the charged black hole (with charge $Q$). For the
Reissner-Nordstrom metric describing such a black hole, one has
$1-2\Phi=(1+\varphi+q \varphi^2)^{-2}(1-q\varphi^2)^2$,
$1+2\Psi=(1+\varphi+q \varphi^2)^{2}$, $q=1-(Q/M)^2$. For this metric,
the expansion (\ref{n2.4}) is also valid with $a={1\over 2}(q-3)$, 
$b={1\over 4}(q+2)$.}
\begin{equation}\label{n2.4} 
\Phi=\stackrel{1}{\phi}+\stackrel{2}{\phi}+\ldots=
\varphi+a\varphi^2+\ldots
\, ,\hspace{0.5cm}
\Psi=\stackrel{1}{\psi}+\stackrel{2}{\psi}+\ldots=
\varphi+b\varphi^2+\ldots
\, .
\end{equation}
The dots denote terms of order $\varphi^3$ and higher and
\begin{equation}\label{n2.5} 
a=-1\, ,\hspace{0.5cm}b={3\over 4}\, .
\end{equation}

A string moving far from the black hole is moving in the perturbed
metric
\begin{equation}\label{n2.6} 
g_{\mu\nu}=\eta_{\mu\nu}+\gamma_{\mu\nu}\, ,
\hspace{0.5cm}\gamma_{\mu\nu}=\stackrel{1}{\gamma}_{\mu\nu}+
\stackrel{2}{\gamma}_{\mu\nu}+\ldots\,\, ,
\end{equation}
\begin{equation}\label{n2.7} 
\stackrel{1}{\gamma}_{\mu\nu}=2\varphi\,\,\delta_{\mu\nu}\, ,
\hspace{0.5cm}
\stackrel{2}{\gamma}_{\mu\nu}=
2 \varphi^2\, \pi_{\mu\nu}\, ,\hspace{0.5cm}
\pi_{\mu\nu}=a\delta^0_{\mu}\delta^0_{\nu}+
b \delta^i_{\mu}\delta^j_{\nu}\delta_{ij}\, .
\end{equation}
Here $i,j=1,2,3$ and $\delta_{ij}$ is the Kronecker $\delta$-symbol.

The perturbation, $\gamma_{\mu\nu}$, of the metric results in the
perturbations $\delta X^{\mu}$ and $\delta h_{AB}$ of the
flat-spacetime solution (\ref{n1.6}) and (\ref{n1.7}). The equations
describing these perturbations can be obtained by perturbing string 
equations (\ref{n1.3}) and (\ref{n1.4}). For this purpose we decompose
the perturbation of the string as
\begin{equation}\label{n2.8} 
\delta X^{\mu}=\chi^m e^{\mu}_{(m)}=
\chi^{R}n^{\mu}_{R}+\chi^{A}X^{\mu}_{A}
 \, ,
\end{equation}
where the four scalar functions of two variables,
$\chi^m(\tau,\sigma)$,  describe the deflection of the string
world-sheet from the plane (\ref{n1.6}). As done earlier, we expand 
$\chi^m={\stackrel{1}{\chi}}{}^m+{\stackrel{2}{\chi}}{}^m+\ldots$ in
powers of $\varphi$. We will also use the expansion of the internal
metric $h_{AB}$
\begin{equation}\label{n2.9} 
h_{AB}=\eta_{AB}+\1{h}_{AB}+\2{h}_{AB}+\ldots\, .
\end{equation}
The first-order corrections will be treated next and then applied to the
general scattering problem. Second-order corrections will be discussed
last to obtain the low-velocity behavior of strings.

\section{First-Order Corrections}

We start by considering  effects which are of the first order in
$\varphi$. In this approximation, the induced metric is
\begin{equation}\label{n3.1} 
{G}_{AB}=\eta_{AB}+\1{\gamma}_{AB}+2\1{\chi}_{(A,B)}\, ,
\end{equation}
where, 
\begin{equation}\label{n3.1a} 
\1{\gamma}_{AB}=\1{\gamma}_{\mu\nu}X^{\mu}_{A}X^{\nu}_{B}=
2\varphi\,\, \mbox{diag}(1-2\sinh^2\beta, 1)\, .
\end{equation}
The perturbation of the constraint equation (\ref{n1.4}) has the form
\begin{equation}\label{n3.2} 
\1{\gamma}_{AB}+2\1{\chi}_{(A,B)}-\1{h}_{AB} -{Q}\eta_{AB}=0\, ,
\end{equation}
where 
\begin{equation}\label{n3.3} 
{Q}={1\over 2}\eta^{CD}\left[\1{h}_{CD}+\1{\gamma}_{CD}+2\1{\chi}_{(C,D)}
\right]\,.
\end{equation}
The tensor $\1{\gamma}_{AB}$ on the two-dimensional world-sheet can be
decomposed as\footnote{In the general case besides the trace and
trace-free ``longitudinal'' part there is also a ``transverse
trace-free'' part $\gamma^{\tiny TT}_{AB}$ which obeys the equation
$\gamma^{\tiny TT}_{AB,C}\eta^{BC}=0$ (see e.g., \cite{GiPe:78}).  It
is easy to verify that a regular solution of this equation on the
two-dimensional world-sheet vanishes.}
\begin{equation}\label{n3.4} 
\1{\gamma}_{AB}={1\over 2}\1{\gamma}\eta_{AB}
+2\xi_{(A,B)}-{1\over 2} \eta_{AB} \eta^{CD}\xi_{C,D} \,.
\end{equation}
By comparing (\ref{n3.2}) and (\ref{n3.4}) one can conclude that one can
always choose $\1{\chi}_{A}$ so that 
\begin{equation}\label{n3.5} 
\1{h}_{AB}=\1{h} \eta_{AB}\, .
\end{equation}
To reach this it is sufficient to put $\1{\chi}_A=-\xi_A$. For this
choice we have
\begin{equation}\label{n3.6} 
\1{\gamma}_{AB}={1\over 2}\1{\gamma} \eta_{AB}
-2\1{\chi}_{(A,B)}-{1\over 2} \eta_{AB} \eta^{CD}\1{\chi}_{C,D} \,
.
\end{equation}
Using (\ref{n3.1a}) we get
\begin{equation}\label{n3.6a} 
\1{\chi}_{0,0}+\1{\chi}_{3,3}=-2\varphi\, \cosh^2\beta\, ,\hspace{0.5cm}
\1{\chi}_{0,3}+\1{\chi}_{3,0}=0\, .
\end{equation}
In what follows we choose these gauge fixing conditions\footnote{  It
should be emphasized that metric $h_{AB}$ on the two-dimensional
world-sheet can always be transformed by means of special choice of the
coordinates to the form $h_{AB}=\Omega\eta_{AB}$. Our choice of the
gauge fixing condition is the infinitesimal form of this relation.}. 

Let us consider now the perturbation of the dynamical equation
(\ref{n1.3}). First, we note that equation (\ref{n3.5}) implies that
$\sqrt{-h}h^{AB}$ is equal to $\eta^{AB}$ up to the terms which are
quadratic in $\varphi$. As a result we have
\begin{equation}\label{n3.7} 
\Box \1{\chi}_m+\eta^{AB}\1{\Gamma}_{\mu,\alpha\beta} X^{\alpha}_{,A}
X^{\beta}_{,B}\, \, e^{\mu}_{(m)}=0\, .
\end{equation}    
In these equations,
\begin{equation}\label{n3.8} 
\Box = -{\partial^2\over \partial\tau^2}+
{\partial^2\over
\partial\sigma^2}\, ,
\end{equation}
and 
\begin{equation}\label{n3.9} 
\1{\Gamma}_{\mu,\alpha\beta}={1\over 2}(\1{\gamma}_{\mu\alpha,\beta}+
\1{\gamma}_{\mu\beta,\alpha}-\1{\gamma}_{\alpha\beta,\mu})
=\varphi_{,\alpha}\delta_{\mu\beta}
+\varphi_{,\beta}\delta_{\mu\alpha}-\varphi_{,\mu}\delta_{\alpha\beta}
 \, .  
\end{equation}  

Since ``longitudinal'' fields $\1{\chi}_A$ are already fixed by our
gauge fixing condition we need to verify that  equations (\ref{n3.7})
for $m=A$ are identically satisfied for this choice and do not give
additional restrictions. For this purpose we remark that
\begin{equation}\label{n3.10} 
\1{\Gamma}_{\mu,\alpha\beta} X^{\alpha}_{,A}
X^{\beta}_{,B}X^{\mu}_{,C}={1\over 2}(\1{\gamma}_{AC,B}+
\1{\gamma}_{BC,A}-\1{\gamma}_{AB,C})\, .  
\end{equation}  
By using this relation and (\ref{n3.6}) it is easy to verify that
(\ref{n3.7}) is satisfied identically for $m=A$.

Let us now discuss dynamical equation (\ref{n3.7}). It  has the form
\begin{equation}\label{n3.11} 
\Box \1{\chi}_R=\left(-{\partial^2\over \partial\tau^2}+ {\partial^2\over
\partial\sigma^2}\right)\1{\chi}_R=\1{f}_R\, ,
\end{equation}
and describes ``transverse'' perturbations of the straight
string under the action of the external gravitational force $\1{f}_R$.
To calculate $\1{f}_R$, note that
\begin{equation}\label{n3.12} 
\1{f}_m=\1{K}_{\mu}e^{\mu}_{(m)}\, ,
\end{equation}
\[
\1{K}_{\mu}=-\eta^{AB}\1{\Gamma}_{\mu,\alpha\beta} X^{\alpha}_{,A}
X^{\beta}_{,B}
\]
\begin{equation}\label{n3.13} 
=\cosh^2\beta \1{\Gamma}_{\mu,00}+
2\sinh\beta\,\cosh\beta \1{\Gamma}_{\mu,01}+
\sinh^2\beta \1{\Gamma}_{\mu,11}- \1{\Gamma}_{\mu,33}
\, .
\end{equation}
Simple calculations give
\begin{equation}\label{n3.14} 
\1{K}_{\mu}=-2\sinh^2\beta \,\,\varphi_{,\mu}+
2\sinh\beta\,\cosh\beta \,\,\varphi_{,1}\delta^0_{\mu}+
2\sinh^2\beta \,\,\varphi_{,1}\delta^1_{\mu}
-2\varphi_{,3}\delta^3_{\mu}
\, .
\end{equation}
Using these results one easily obtains
\begin{equation}\label{n3.15} 
\1{f}_1=2\sinh^2\beta\, \cosh\beta \, \, \varphi_{,X}\, ,
\hspace{0.5cm}
\1{f}_2=-2\sinh^2\beta\,\, \varphi_{,Y}
\, ,
\end{equation}
\begin{equation}\label{n3.16} 
\1{f}_0=2\sinh\beta\, \cosh^2\beta \, \, \varphi_{,X}\, ,
\hspace{0.5cm}
\1{f}_3=-2\cosh^2\beta\,\, \varphi_{,Z}
\, ,
\end{equation}

The components $f_R$ normal to the string
world-sheet are components of the physical force acting on the
string. The components $f_A$ acting along the string provided a motion
along the world-sheet which has no physical meaning and can be removed
by coordinate transformations.

\section{String Scattering}

Equation (\ref{n3.11}) for string propagation in a weak
gravitational field can be easily solved. The retarded Green's function
for the 2D $\Box$-operator is
\begin{equation}\label{n4.1} 
{G}_{0}\left(\sigma,\tau \mid \sigma',\tau'\right) = {1 \over 2}\,
\Theta(\tau - \tau' - \mid \sigma - \sigma' \mid)\, .
\end{equation}
Using this Green's function we can write a solution of (\ref{n3.11}) in
the form\footnote{In this section we consider only first-order
corrections to string motion. The superscript 1 in $\1{\chi}_m$ and
similar quantities in this section is omitted for briefness.}
\begin{equation}\label{n4.2} 
{\chi}_m(\tau,\sigma) ={\chi}_m^0(\tau,\sigma)+{\chi}_m^+(\tau+\sigma)+
{\chi}_m^-(\tau-\sigma)\, ,
\end{equation}
where
\[
{\chi}_m^0(\tau,\sigma)=-\int_{{\tau}_{0}}^{{\tau}}{d\tau'}
\int_{-\infty}^{\infty}{d\sigma'}
{G}_{0}\left(\sigma,\tau \mid \sigma',\tau'\right) 
{f}_m(\tau',\sigma')
\]
\begin{equation}\label{n4.3} 
=-{1\over 2}\int_{\tau_0}^{\tau} d\tau'
\int_{\sigma-(\tau-\tau')}^{\sigma+\tau-\tau'} d\sigma'
f_m(\tau',\sigma')
\, 
\end{equation}
is a solution of inhomogeneous equation and ${\chi}_m^{\pm}$ are
solutions of homogeneous equation which are fixed by the initial data 
\begin{equation}\label{n4.4} 
{\chi}_m(\tau_0,\sigma)={\chi}_m^+(\tau_0+\sigma)+
{\chi}_m^-(\tau_0-\sigma)\, ,\hspace{0.5cm}
\dot{{\chi}}_m(\tau_0,\sigma)=\dot{\chi}_m^+(\tau_0+\sigma)+
\dot{\chi}_m^-(\tau_0-\sigma)\, .
\end{equation}

Let us first consider perturbations perpendicular to the direction of
motion (the $Y$-direction), described by $\chi_2$. We assume that
initially (at the infinite past) $\chi_2=0$. For this initial condition
at $\tau_0=-\infty$, \  ${\chi}_2^{\pm}=0$. The asymptotic final
solution (at the infinite future) takes the form
\begin{equation}\label{n4.5} 
{\chi}_2(\tau=\infty)=
\lim_{\tau\rightarrow \infty}{\chi}_2^0(\tau,\sigma)
=-{1\over 2}\int_{-\infty}^{\infty}{d\tau'}
\int_{-\infty}^{\infty}{d\sigma'}\, \, 
{f}_2(\tau',\sigma')\, .
\end{equation}
Substituting expression (\ref{n3.15}) for $f_2$ and making a change of
variables of integration from $(\tau,\sigma)$ to $(X,Z)$ we get
\begin{equation}\label{n4.6} 
{\chi}_2(\tau=\infty)=-\sinh\beta\, \int_{-\infty}^{\infty}{dX}
\int_{-\infty}^{\infty}{dZ}\, \, {\partial\varphi\over \partial Y}(X,Y_0,Z)\,
.
\end{equation}
The integral represents the flux of the Newton gravitational field
through the plane $Y=Y_0$ which is equal to $2\pi M$ (that is a half of
the total flux $4\pi M$). Using this simple observation we get that as
the result of scattering the string as the whole is displaced in
$Y$-direction by a constant value
\begin{equation}\label{n4.7} 
{\chi}_2(\tau=\infty)=-2\pi M\sinh\beta\, .
\end{equation}

At late but finite time only part of the string is displaced.  The size of the
displaced region grows with the velocity of light.  The transition between the
``old'' and ``new'' phases occurs at two kinks moving in the opposite
direction. The late time solution can be found explicitly, and is schematically
shown in Figure~1.  The background world-sheet sweeps out a flat plane in
space; denote this the {\it in-string plane}, the plane in which the motion of
the string lies at early times. At late times, the scattered string approaches
another plane, offset from the in-string plane down by $|{\chi}_{2}^{\infty}|$;
denote this the {\it out-string plane}. As the energy acquired by the string is
propagated to infinity through the two kinks, more and more of the string falls
to the out-string plane. The asymptotic deflection, ${\chi}_{2}^{\infty}$, is
determined by the properties of the encounter, and is given by (\ref{n4.7}).

\begin{figure}[3Dpert]
\let\picnaturalsize=N
\def\picsize{6cm}
\def\picfilename{3D_scatter.eps}
\ifx\nopictures Y\else{\ifx\epsfloaded Y\else\input epsf \fi
\let\epsfloaded=Y
\centerline{\ifx\picnaturalsize N\epsfxsize \picsize\fi
\epsfbox{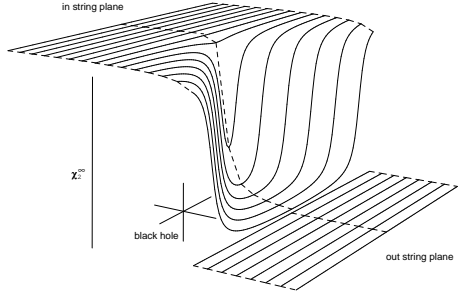}}}\fi
\caption{Weak-field scattering. }
\end{figure}
 
If a straight string starts its motion ($\tau=0$) at $X_0<0$, 
\begin{equation}\label{n4.8} 
{\chi}_2(0,\sigma)=\dot{\chi}_2(0,\sigma)=0\, ,
\end{equation}
and the solution has the form
\begin{equation}\label{n4.9} 
{\chi}_2(\tau,\sigma)=-M\sinh\beta\,\left[H_{+}(\tau,\sigma)+
H_{-}(\tau,\sigma)\right]
\, ,
\end{equation}
where
\[
H_{\pm}(\tau,\sigma)=
\arctan{\left[{{Y}_{0}^{2} + ({X}_{0}+\tau\,\sinh{\beta})\,
({X}_{0}+{s}_{\pm}\,\sinh{\beta}) \over 
 {Y}_{0}  \,\sinh \beta \, \, R(\tau,\sigma)}\right]} 
\]
\begin{equation}\label{n4.10} 
-\arctan{\left[{{X}_{0}\left({X}_{0} + {s}_{\pm} \, \sinh{\beta}\right) +
{Y}_{0}^{2} \over 
 {Y}_{0}  \,\sinh{ \beta }\sqrt{{\rho}^{2}+{s}_{\pm}^{2}}}\right]}\, .
\end{equation}
We use the notation
\begin{equation}\label{n4.11} 
R^{2}(\tau,\sigma) =
{\left({X}_{0}+\tau \sinh{\beta}\right)}^{2} + {Y}^{2}_{0} +
{\sigma}^{2}\, ,\hspace{0.5cm}
{\rho}^{2}  = {X}_{0}^{2} + {Y}_{0}^{2}\, ,\hspace{0.5cm}
s_{\pm}=\tau\pm\sigma\, .
\end{equation}

At the moment when $\tau\sinh\beta=-X_0$, the string passes at the
closest distance from the source of the gravitational field. In order to
study the late time behavior of the string, let us consider the limit
when $X_0=-L$, $\tau\sinh\beta=2L$, and $L\rightarrow\infty$. In this
limit, the expression for $H_{\pm}$ simplifies to
\begin{equation}\label{n4.12} 
H_{\pm}\approx\arctan{\left[{L\pm\sigma\sinh\beta\over{Y_0\sinh\beta
\sqrt{1+(\sigma/L)^2}}}\right]}+
\arctan{\left[{L\pm\sigma\sinh\beta\over{Y_0\sinh\beta
\sqrt{1+((2/\sinh\beta)\pm\sigma/L)^2}}}\right]}\, .
\end{equation}
For $H_{\pm}$ the kink is located near $\sigma=\mp L/\sinh\beta$. Using
this fact we can further simplify the asymptotic expression for $H_{\pm}$
and to write it in the form
\begin{equation}\label{n4.13} 
H_{\pm}\approx 2\arctan{\left[{L\pm\sigma\sinh\beta
\over{Y_0\cosh\beta}}\right]}\,
.
\end{equation}

At late time in the asymptotic region where
$g_{\mu\nu}\approx\eta_{\mu\nu}$ the action (\ref{n1.1}) can be written
as the sum of the action for the straight string and a term which is
quadratic in perturbations. This term is of the form (for details see
Ref.\cite{FrLa:94})
\begin{equation}\label{n4.12a} 
I_2=-{\mu \over 2}\,\int d\tau d\sigma \sqrt{-h}h^{AB}\chi^R_{,A}\chi^R_{,B}\, .
\end{equation}
Hence the contributions of $\chi_2$ to the energy is
\begin{equation}\label{n4.12b} 
E = {\mu \over 2} \int_{-\infty}^{\infty} {
d\sigma\,\left\{ {\left({\partial \chi_2 \over \partial \tau}\right)}^{2}+
{\left({\partial \chi_2 \over \partial \sigma}\right)}^{2}\right\}}\, .
\end{equation}
Using solution (\ref{n4.13}) and
\begin{equation}\label{pert.25a}
{\partial \chi_2 \over \partial \tau} = 
{\partial \chi_2 \over \partial L}\,{\partial L \over \partial \tau} =
{\sinh{\beta} \over 2}{\partial \chi_2 \over \partial L}\, ,
\end{equation}
the integrals can be evaluated in a straightforward manner. In the limit
$L~\rightarrow~\infty$, the energy carried away by each of the kinks
has a very simple form
\begin{equation}\label{pert.25b}
E = {5 \mu \over 32 \pi} {{A}_{\infty}^{2} \over w} \, ,
\end{equation}
where $w = Y_0\,\coth{\beta}$ is the width of the kinks, and 
${A}_{\infty} = \mid \chi_2(\tau = \infty)\mid$ their late-time
amplitude.

One can also obtain solutions for the other components of $\chi_m$.
Substituting (\ref{n3.15}) and (\ref{n3.16}) into (\ref{n4.3}) and
performing the integrations one gets
\begin{eqnarray}\label{pert.21}
\chi_0(\tau,\sigma) & = &  M\,{\cosh}(\beta)
\left[
\ln{\left({F}_{+}(\tau,\sigma)\right)} + 
\ln{\left({F}_{-}(\tau,\sigma)\right)}
\right. \\
\nonumber &  + & {1 \over 2}\,{\cosh}(\beta)\,
\left. \left[
\it{sgn}{({s}_{+})} \ln{\left({G}_{+}(\tau,\sigma)\right)} + 
\it{sgn}{({s}_{-})} \ln{\left({G}_{-}(\tau,\sigma)\right)}
\right]\right]\\
{\chi}_{1}(\tau,\sigma) & = &  -M \,{\sinh}(\beta)
\left[
\ln{\left({F}_{+}(\tau,\sigma)\right)} + 
\ln{\left({F}_{-}(\tau,\sigma)\right)}
\right. \\
\nonumber &  + & {1 \over 2}\,{\cosh}(\beta)\,
\left.\left[
\it{sgn}{({s}_{+})} \ln{\left({G}_{+}(\tau,\sigma)\right)} + 
\it{sgn}{({s}_{-})} \ln{\left({G}_{-}(\tau,\sigma)\right)}
\right]\right]\\
{\chi}_{3}(\tau,\sigma) & = & M\,{\cosh}(\beta)\,
\left[
\ln{\left({F}_{+}(\tau,\sigma)\right)} - 
\ln{\left({F}_{-}(\tau,\sigma)\right)}
\right]
\end{eqnarray}
where
\begin{eqnarray}\label{pert.23}
{F}_{\pm}(\tau,\sigma) & = & 
{R\,\cosh{\beta} + \tau\,{\cosh}^{2}{\beta} + {X}_{0}\,\sinh{\beta} - 
{s}_{\pm} 
\over 
\cosh{\beta}\,\sqrt{{\rho}^{2} + {s}_{\pm}^{2}} + {X}_{0}\,\sinh{\beta}
- 
{s}_{\pm}}\\
{G}_{\pm}(\tau,\sigma) & = & 
{\sqrt{{\rho}^{2} + {s}_{\pm}^{2}} - \mid {s}_{\pm} \mid 
\over 
\sqrt{{\rho}^{2} + {s}_{\pm}^{2}} + \mid {s}_{\pm} \mid }
\end{eqnarray}

As was done for $\chi_2$, expressions (\ref{pert.23}) can be rewritten
in terms of the parameter $L$ (with $L \gg Y_0$),
\begin{eqnarray}\label{pert.23a}
{F}_{\pm} & = & 
{\cosh{\beta} \sqrt{1+(\sigma/L)^2} + 
\left({\cosh}^{2}{\beta} + 1\right)/\sinh{\beta} - (2/\sinh\beta\pm\sigma/L)
\over 
\cosh{\beta}\,\sqrt{1+(2/\sinh\beta\pm\sigma/L)^2} - \sinh{\beta}- 
(2/\sinh\beta\pm\sigma/L)}\\
{G}_{\pm} & = & 
{\sqrt{1+(2/\sinh\beta\pm\sigma/L)^2} - \mid 2/\sinh\beta\pm\sigma/L \mid 
\over 
\sqrt{1+(2/\sinh\beta\pm\sigma/L)^2} + \mid 2/\sinh\beta\pm\sigma/L \mid}
\end{eqnarray}

In rewriting the expressions in terms of the location of the kink, 
$\sigma=\mp L/\sinh\beta$, one sees that, ${F}_{\pm} \rightarrow
\infty$ and ${G}_{\pm} \rightarrow \left(\cosh{\beta}-1\right)/
\left(\cosh{\beta}+1\right)$. Whereas the contributions $\ln G_\pm$ are
well behaved, those from $\ln F_\pm$ generate a logarithmic divergence
in $\chi_0$ and $\chi_1$. This divergence is the result of the long-range
nature of gravitational forces and it is similar to the logarithmic
divergence of the phase for the Coulomb scattering in quantum mechanics.
It vanishes for potentials vanishing at infinity rapidly enough. 

The perturbations $\chi_m$ are illustrated by Figures~2--4, where the solutions
are applied to the case of a straight string with initial velocity $v =
\tanh{\beta} = 0.76c$ and impact parameter  $b = {Y}_{0} = 40 {r}_{g}$  (for
$-2000 {r}_{g} < \sigma < 2000 {r}_{g}$). These figures  show each perturbation
at late proper time, when the string is well past the black hole. 
The ${\chi}^{0}$ and ${\chi}^{1}$ perturbations (Figure~2)
are exceedingly small.  The ${\chi}^{2}$ perturbation (Figure~3) describes the
deformation of the string normal to the background world-sheet; the two
kink-like pulses propagating away from the $Z = 0$ plane at the speed of light
are clearly visible, and their amplitude is considerably larger that any of the
other perturbations.  These pulses carry energy away to infinity and, in the
process, shift the string's late-time position roughly $3.5 {r}_{g}$ below the
original position. The ${\chi}^{3}$  perturbation (Figure~4) represents lateral
displacements of points on the string towards the $Z = 0$ plane; the amplitude
of these displacements is small in the weak field limit, but will become
significant in the limit of ultra-relativistic velocity and shallow impact
parameter, where lateral  displacements of the string are involved in transient
loop formation \cite{DVFr:98}.

\begin{figure}[Tpert]
\let\picnaturalsize=N
\def\picsize{7cm}
\def\picfilename{chi0_chi1.eps}
\ifx\nopictures Y\else{\ifx\epsfloaded Y\else\input epsf \fi
\let\epsfloaded=Y
\centerline{\ifx\picnaturalsize N\epsfxsize \picsize\fi
\epsfbox{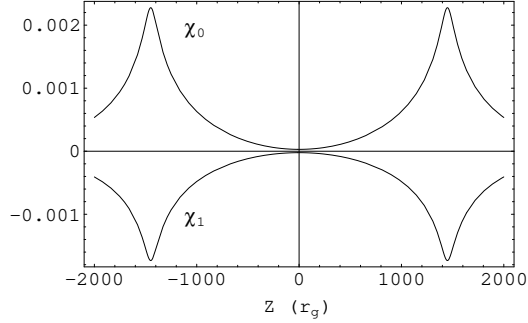}}}\fi
\caption{$\chi_0$ and $\chi_1$ perturbations.}
\end{figure}

\begin{figure}[Ypert]
\let\picnaturalsize=N
\def\picsize{7cm}
\def\picfilename{chi2.eps}
\ifx\nopictures Y\else{\ifx\epsfloaded Y\else\input epsf \fi
\let\epsfloaded=Y
\centerline{\ifx\picnaturalsize N\epsfxsize \picsize\fi
\epsfbox{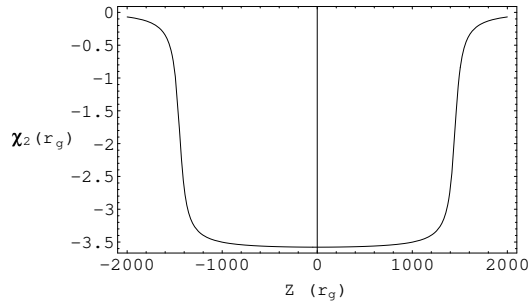}}}\fi
\caption{$\chi_2$ perturbation.}
\end{figure}

\begin{figure}[Zpert]
\let\picnaturalsize=N
\def\picsize{7cm}
\def\picfilename{chi3.eps}
\ifx\nopictures Y\else{\ifx\epsfloaded Y\else\input epsf \fi
\let\epsfloaded=Y
\centerline{\ifx\picnaturalsize N\epsfxsize \picsize\fi
\epsfbox{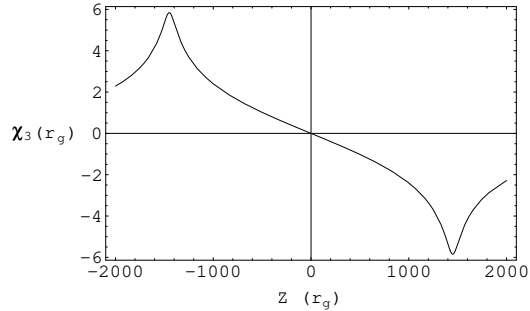}}}\fi
\caption{$\chi_3$ perturbation.}
\end{figure}

The perturbation solutions can be used to reconstruct the full world-sheet of
the string in Cartesian coordinates, using 
\begin{equation}\label{pert.24}  
{\cal X}^{\mu}=  {X}^{\mu} + \chi^m e^{\mu}_{(m)} \, .  
\end{equation}  
Such a reconstruction is shown in Figure~5, and was also used to generate the
schematic representation in Figure~1.  Figure 5 shows a  sequence of string
configurations separated by constant intervals of proper time in two separate
views. The view on the left looks down on the XZ plane and shows the outward
propagation of the two pulses (the black hole lies at the origin). Note that the
view is a 3D projection; the kinks appear to extend in the X-direction, 
but they actually lie in the Y-direction (the effect is
an artifact of the viewpoint chosen for this view). The view on
the right looks toward the origin along the direction of motion, and shows the
growth of the perturbations along the Y axis. Comparing to
Figures 2 through 4, it is easily seen that the shape
of the perturbed world-sheet is almost completely determined by the $\chi_2$
perturbation. The contribution from the other perturbations is undetectable on
the scale used in Figure 5.  At late times, the string is deflected by an
amount $|{\chi}_{2}^{\infty}|$, as given by (\ref{n4.7}).

\begin{figure}[Zpert]
\let\picnaturalsize=Y
\def\picsize{7cm}
\def\picfilename{chi_composite.eps}
\ifx\nopictures Y\else{\ifx\epsfloaded Y\else\input epsf \fi
\let\epsfloaded=Y
\centerline{\ifx\picnaturalsize N\epsfxsize \picsize\fi
\epsfbox{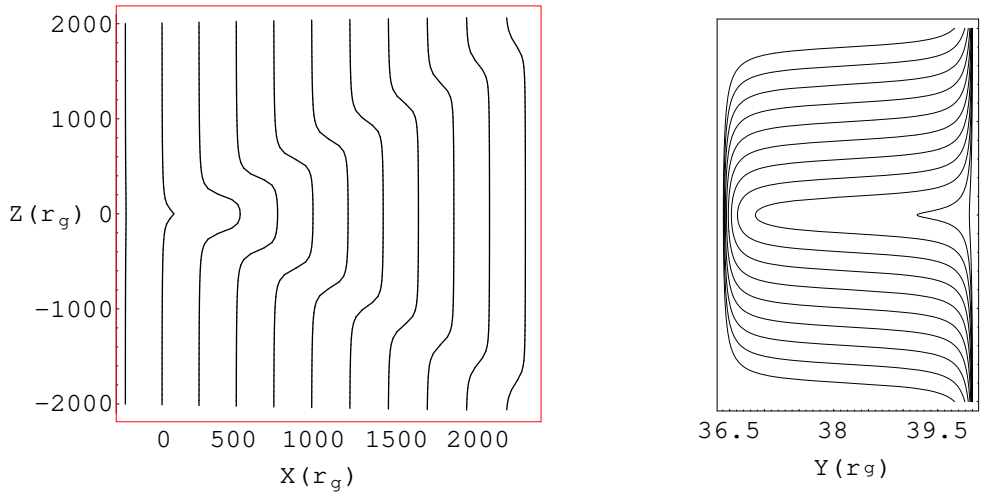}}}\fi
\caption{Reconstruction of perturbed string (Cartesian coordinates).}
\end{figure}

To summarize, the scattering of the string in the weak gravitational
fields calculated in the first order in $\varphi$ 
results in the displacement of the string in the direction
perpendicular to the motion by the value ${\chi}_2(\tau=\infty)=-2\pi
GM\sinh\beta/c^2$. At any finite but large value of $\tau$ a solution
represents a kink and anti-kink of the width $Y_0\coth\beta=Y_0 c/v$
propagating in the opposite directions with the velocity of light and
leaving behind them the string in the new ``phase'' with
$Y=Y_0+{\chi}_2(\tau=\infty)$.

\section{Low-Velocity Limit}

As was already mentioned, the components $f_R$ normal to the string
world-sheet are components of the physical force acting on the
string.  As can be seen from (\ref{n3.15}), the force $f_R$ acting on
the string vanishes in the limit $v\rightarrow 0$. This fact has a
simple physical explanation. As was shown in \cite{FrSkZeHe:89}, a
static string configuration in a static spacetime is a geodesic in a
spacetime with the metric $|g_{00}|g_{ij}$, which in our case takes
the form
\begin{equation}\label{n5.1} 
dS^2=|g_{00}|g_{ij}dx^i dx^j=(1-2\Phi)(1+2\Psi)(dX^2+dY^2+dZ^2)\, .
\end{equation}
In the leading order, $\Phi=\Psi=\varphi$, and the string is a straight
line \cite{ShVi:94}. In other words, in the first-order approximation a
force acting on a static string in a static spacetime of a black hole
vanishes. For this reason, the leading terms in the expansion of the
force are of the second order in $\varphi$, and they remain so
until $v/c\sim (GM/b_{imp})^{1/2}$. In this section we discuss the effect of
these second order terms on the motion of the string in the limit of
very small velocities.

Substituting (\ref{n2.8}) into the dynamical equations (\ref{n1.3}) we
get
\begin{equation}\label{n5.2} 
n^{\mu}_R\,\, \Box \2{\chi}{}^{\! R}=
\2{f}{}^{\! \mu}\equiv A^{\mu}+B^{\mu}\, ,
\end{equation}
where the $\Box$-operator is given by (\ref{n3.10}) and,
\begin{equation}\label{n5.3} 
A^{\mu}=-\eta^{AB}\2{\Gamma}{}^{\!
\mu}_{\alpha\beta}X^{\alpha}_{,A} X^{\beta}_{,B}\, ,
\hspace{0.5cm}
B^{\mu}=-\eta^{AB}\1{\Gamma}{}^{\!
\mu}_{\alpha\beta}(\1{\chi}{}^{\! m}_{,B} X^{\alpha}_{,A}
e^{\beta}_{(m)}+\1{\chi}{}^{\! m}_{,A} X^{\beta}_{,B}
e^{\alpha}_{(m)}) \, .
\end{equation}
Note that
\begin{equation}\label{n5.4} 
\2{\Gamma}{}^{\!\mu}_{\alpha\beta}=\eta^{\mu\nu}
\2{\Gamma}_{\nu,\alpha\beta}+
\1{\gamma}{}^{\!\mu\nu}\1{\Gamma}_{\nu,\alpha\beta}\,
\end{equation}
where $\1{\Gamma}_{\nu,\alpha\beta}$ is given by (\ref{n3.9}) and,
\begin{equation}\label{n5.5} 
\2{\Gamma}_{\nu,\alpha\beta}=2\varphi\,
(\varphi_{,\alpha}\, \pi_{\nu\beta}
+\varphi_{,\beta}\, \pi_{\nu\alpha}-\varphi_{,\nu}\, \pi_{\alpha\beta})
 \, .  
\end{equation}

We focus our attention on the corrections to the motion of the string in
$Y$-direction. It is easy to verify that
\begin{equation}\label{n5.6} 
A^{2}=\cosh^2\beta \2{\Gamma}{}^{\! 2}_{00}- \2{\Gamma}{}^{\! 2}_{33}=
2\varphi\varphi_{,2}[(1-a)\cosh^2\beta -(1-b)]\, .
\end{equation}
Calculations also give
\begin{equation}\label{n5.7} 
B^2=2\varphi_{,2}\left[(\cosh^2\beta+\sinh^2\beta)\1{\chi}_{0,0}-
\sinh(2\beta)\1{\chi}_{1,0}+\1{\chi}_{3,3}\right]-
2\varphi_{,3}\1{\chi}_{2,3}\, .
\end{equation}
At low velocities one has
\begin{equation}\label{n5.8} 
A^{2}\sim 2\varphi\varphi_{,2} (b-a)\, ,\hspace{0.5cm}
B^2\sim 2\varphi_{,2}
(\1{\chi}_{0,0}+\1{\chi}_{3,3})-2\varphi_{,3}\1{\chi}_{2,3}\, .
\end{equation}
Equation (\ref{n4.9}) shows that $\1{\chi}_{2}$ vanishes at
$\beta\rightarrow 0$.  Using  relation (\ref{n3.6a}), we finally get
\begin{equation}\label{n5.9} 
\2{f}_{2}\sim -2(2+a-b) \varphi\varphi_{,2} \, .
\end{equation}

Using a relation similar to (\ref{n4.5}) we get
\begin{equation}\label{n5.10} 
\2{\chi}_2(\tau=\infty)={2+a-b\over 2\sinh\beta}\, 
\int_{-\infty}^{\infty}{dX}
\int_{-\infty}^{\infty}{dZ}\, \, {\partial\varphi^2\over \partial
Y}(X,Y_0,Z)\,
.
\end{equation}
Calculating the integral one gets
\begin{equation}\label{n5.11} 
\2{\chi}_2(\tau=\infty)=-{\pi M^2(2+a-b)\over Y_0\sinh\beta }\,.
\end{equation}

For the scattering of the string on the Schwarzschild black hole, $a=-1$ and
$b=3/4$, so that one has
\begin{equation}\label{n5.12} 
\2{\chi}_2(\tau=\infty)=-{\pi (GM)^2\over 4c^4 Y_0\sinh\beta }\,.
\end{equation}
Using the expressions for the coefficients $a$ and $b$ for the
scattering on a Reissner-Nordstrom black hole (see footnote 1) one gets
\begin{equation}\label{n5.13} 
\2{\chi}_2(\tau=\infty)=-{\pi ((GM)^2-GQ^2)\over 4c^4 Y_0 \sinh\beta }\,.
\end{equation}
These results are in the complete agreement with the results obtained by
Page \cite{Page:98}.

\section{Conclusions}

We analyzed the motion of a cosmic string in the gravitational field of
a black hole in the approximation where the field is weak. In
particular, we demonstrated that after passing the black hole, the
string continues its motion with the same velocity $\bf v$ as before
scattering, but it is displaced in the direction to the black hole and
perpendicular to $\bf v$ by the distance $\Delta b\sim -2\pi
GM\sinh\beta/c^2 -\pi (GM)^2/ (4c^3 v b_{imp})$, where $b_{imp}$ is the
impact parameter. If $|b_{imp}-\Delta b|\gg r_g=2GM/c^2$ the string
moves always in a weak field. If $|b_{imp}-\Delta b|\sim r_g=2GM/c^2$
the string enters the strong-field region near the black hole and it
can be captured. This allows us to give the following estimate of the
critical capture impact parameter. For low velocity, $(v/c)\ll
1$,
\begin{equation}\label{n6.1}  
b_{capture} \sim {GM \over c^2} 
\left(\pi\over 4\right)^{1/2}\,\left(v\over c\right)^{-1/2}\,  .
\end{equation}
For $v/c>(GM/b_{imp})^{1/2}$ the first term in $\Delta b$ dominates.
This means that the capture impact parameter grows for both small and
large values of $v/c$, and hence there exists a velocity $v$ for which
the capture impact parameter has minimum value. 
This conclusion is
confirmed by the results of the numerical computations of the capture
impact parameter \cite{DVFr:98}. The numerical results demonstrate also
that for the ultrarelativistic velocities the critical impact parameter
$b_{capture}$ reaches the value $3\sqrt{3}GM/c^2$, that is the same
value as the capture parameter for the ultrarelativistic particles.

Besides giving us a qualitative understanding of the scattering and capture of
cosmic strings by black holes, the analysis of the weak-field approximation is
important for the numerical study of these processes in the strong field.  
Before and after the scattering the string moves far from the black hole, where
the gravitational field is weak. Thus one can use the above analysis to
provide a well-defined description of ``in''- and ``out''-states of the string
and to formulate the scattering problem. We discuss this in Ref.\cite{DVFr:98}.

\vspace{12pt} 
{\bf Acknowledgments}:\ \ This work was partly supported by the Natural
Sciences and Engineering Research Council of Canada.  One of the
authors (V.F.) is grateful to the Killam Trust for its  financial
support. The authors are grateful to Arne Larsen for early insights
into the perturbative equations of motion. The authors are also
grateful to Don Page,  who made  his paper \cite{Page:98} available
prior to publication. We also  thank him for various discussions.

\end{document}